\documentclass[%
 aip,
 amsmath,amssymb,
 reprint,%
]{revtex4-1}

\usepackage[textsize = scriptsize, color = pink]{todonotes}
\usepackage{color}
\usepackage{float}
\usepackage{graphicx}
\usepackage{amsmath,amstext,amssymb,amsthm}
\usepackage{siunitx}
\usepackage[normalem]{ulem}
\DeclareSIUnit{\nothing}{\relax}
\PassOptionsToPackage{bookmarks={false}}{hyperref}
\usepackage[hidelinks]{hyperref}
\usepackage{svg}

\usepackage[normalem]{ulem}
\usepackage{upgreek}
\usepackage{threeparttable}
\usepackage{xspace}
\usepackage{comment}
\usepackage{setspace}
\usepackage{xcolor}
\definecolor{orange}{rgb}{0.59, 0.29, 0.1}
\definecolor{yellow}{rgb}{1.0, 0.7, 0.0}

\draft 

\begin{document}


\title{Coherent Phonon-Driven Band Renormalizations in 1\textit{T}$'$-MoTe$_2$} 



\author{Carl E.~Jensen}
\email{cjensen@physik.uni-kiel.de}
\affiliation
{Institute of Experimental and Applied Physics, Kiel University, 24118 Kiel, Germany
}
\author{Christoph~Emeis}
\affiliation{Institute of Theoretical Physics and Astrophysics, Kiel University, 24118 Kiel, Germany
}
\author{Stephan~Jauernik}
\affiliation{Institute of Experimental and Applied Physics, Kiel University, 24118 Kiel, Germany
}
\author{Petra~Hein}
\affiliation{Institute of Experimental and Applied Physics, Kiel University, 24118 Kiel, Germany
}
\author{Fabio~Caruso}
\affiliation{Institute of Theoretical Physics and Astrophysics, Kiel University, 24118 Kiel, Germany
}
\author{Michael~Bauer}
\homepage{https://www.physik.uni-kiel.de/de/institute/ag-bauer}
\affiliation{Institute of Experimental and Applied Physics, Kiel University, 24118 Kiel, Germany
}
\affiliation{
Kiel Nano, Surface and Interface Science KiNSIS, Kiel University, 24118 Kiel, Germany
}


\date{\today}

\begin{abstract}
Here, we investigate phonon mode- and electron band-selective electron–phonon couplings in centrosymmetric 1\textit{T}$'$-MoTe$_2$ using time- and angle-resolved photoemission spectroscopy combined with frequency-domain analysis. Femtosecond near-infrared pulses excite coherent $A_g$-symmetric phonon modes at 2.34 THz, 3.34 THz, and 3.86 THz, which manifest as oscillatory modulations in photoemission intensity and binding energy across the valence bands. Pixel-wise Fourier analysis using recently developed methodologies reveals pronounced band selectivity with distinct coupling strengths for different electronic states and phonon modes, enabling the evaluation of band-renormalization amplitudes in the range of few meV. \textit{Ab initio} calculations qualitatively reproduce the experimentally observed coupling patterns and relative trends, demonstrating the capability of combined experimental and theoretical approaches to resolve ultrafast electron–phonon interactions in quantum materials.
\end{abstract}


\maketitle 


\section*{Introduction}
Transition metal dichalcogenides (TMDCs) have emerged as a versatile platform for exploring novel quantum phenomena due to their rich polymorphism and tunable electronic properties~\cite{Manzeli2017, Chhowalla2013}. Among the TMDCs, molybdenum ditelluride (MoTe$_2$) is of particular interest, as it hosts a variety of crystal phases with distinct topological and electronic characteristics. These include semiconducting, semimetallic, Weyl-semimetallic, and superconducting phases~\cite{Qi2016, Deng2016}, making MoTe$_2$ a promising candidate for future quantum electronic and optoelectronic applications.

In this material, coherent phonons and their coupling to electronic degrees of freedom are of particular interest, as they can drive structural phase transitions and thus enable ultrafast control of the associated electronic properties through optical excitation. Examples include a transient phase transition from the topologically non-trivial \textit{Td} phase to the topologically trivial 1\textit{T’} phase of MoTe$_2$ through coherent excitation of a low-frequency shear mode~\cite{Zhang2019}, and the prediction of a phase transition from the semiconducting 2\textit{H}-phase to the semimetallic 1\textit{T’} phase of monolayer MoTe$_2$ through a complex interplay of different coherently excited phonon modes~\cite{Guan2022}.

Time- and angle-resolved photoemission spectroscopy (tr-ARPES) is a powerful tool for directly investigating the ultrafast dynamics of electronic excitations and their coupling to phonons~\cite{Perfetti2007, Johannsen2013, Stange2015}. Furthermore, tr-ARPES allows the observation of changes in the electronic band structure due to coupling to coherent phonons~\cite{Perfetti2006}. A phonon frequency-selective analysis of these data (frequency domain ARPES – FDARPES) enables a mode-specific projection of the electron-phonon interaction onto the measured electronic band structure~\cite{hein2020mode, DeGiovannini2020}.

In this work, we employ tr-ARPES and FDARPES to investigate phonon mode and electron band-resolved electron--phonon interactions in 1\textit{T}$'$-MoTe$_2$. A near infrared femtosecond laser pulse launches the excitation of several coherent phonon modes, whose signatures are observed as oscillatory modulations in the photoemission intensity and binding energy of electronic bands. Our FDARPES analysis reveals a pronounced selectivity in the coupling between specific bands and the A$_g$-symmetric phonon modes at $\nu_1=\SI{2.34}{THz}$, $\nu_2=\SI{3.34}{THz}$, and $\nu_3=\SI{3.86}{THz}$. In a recent study of the related compound WTe$_2$ in the $Td$ phase an analysis method was presented that makes it possible to disentangle the various causes of such oscillations and quantify their amplitudes ~\cite{gauthier2025analysis}. Application of this method to our photoemission data allows us to quantitatively extract amplitudes of observed band renormalizations that arise from changes in the wave function overlap and the screening behavior within the crystal lattice as the electronic potential landscape is modulated by the periodic ionic displacements.

In addition, we performed \textit{ab initio} calculations of coherent phonon-driven band renormalizations in 1\textit{T}$'$-MoTe$_2$, which we analyze in analogy to the experimental data. In a direct comparison, we find good qualitative agreement between experiment and theory. Possible causes for observed quantitative deviations include a potential systematic underestimation of excitation amplitudes by the evaluation method applied to the experimental data and uncertainties regarding the photoexcitation density.

\section*{Methods}
\label{sec:methods}

1\textit{T'}-$\mathrm{MoTe}_2$ single crystals (2D Semiconductors), were mounted on a tantalum sample holder using a two-component epoxy. A cleaving post was affixed to the crystal surface with the same epoxy, and the sample was subsequently cleaved in ultrahigh vacuum ($\leq 2\cdot10^{-10}$ mbar). Tr-ARPES measurements were performed using a pump–probe scheme as schematically illustrated in Fig.~\ref{fig1}(a). Near-infrared (NIR) p-polarized pump pulses at \SI{837}{\nano m} (\SI{1.5}{eV}) were used for excitation at an absorbed fluence of \(\approx~\SI{0.3}{\milli J/\centi m^2}\).
The response of the electronic system to the photoexcitation was probed using time-delayed and s-polarized near-ultraviolet (NUV) pulses at \SI{210}{\nano m} (\SI{5.9}{eV}). The time resolution was determined to be \SI{130}{\femto s} corresponding to the full width at half maximum of the NIR-NUV cross-correlation signal. The emitted photoelectrons were detected using a hemispherical electron energy analyzer. All experiments were performed at room temperature. More details about the experimental setup can be found in \cite{jauernik2018probing}.\\
The electronic band structure, phonon dispersions, and electron-phonon coupling (EPC) matrix elements were determined using density functional theory (DFT) and density functional perturbation theory (DFPT). The time-dependent electronic distribution function $f_{n\mathbf{k}}(t)$ and the coherent lattice displacements $Q_{\nu}(t)$ are obtained by solving the time-dependent Boltzmann equations and the coherent-phonon equation of motion. Our first principles simulations were conducted using the open-source computer codes {\tt Quantum ESPRESSO} \cite{giannozzi2009quantum,giannozzi2017advanced} and {\tt EPW} \cite{lee2023electron}. Further information on the theory and the computational details can be found in \cite{emeis2025coherent} and in the supplementary information \cite{Supplemental}, respectively.

\section*{Results and Discussion}
\begin{figure}
        \includegraphics[width=\linewidth]{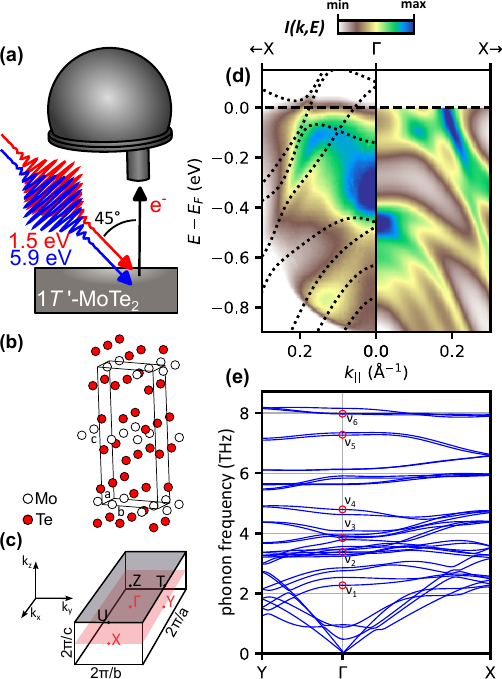}
        \setlength{\abovecaptionskip}{-10pt}
        \caption{(a) Schematic illustration of the tr-ARPES experiment. (b) Crystal structure of 1\textit{T'}-$\mathrm{MoTe}_2$. (c) First Brillouin zone of 1\textit{T'}-$\mathrm{MoTe}_2$ with high symmetry points indicated. (d) Room-temperature ARPES spectrum at \(h\nu=\SI{5.9}{eV}\) (left panel)  and  spectral function calculated from the DFT band structure (right panel) along the \(\Gamma\)-X direction. The experimental data is overlayed with the DFT band structure (dashed lines). $k_{\parallel}$ denotes the electron wave vector parallel to the sample surface. (e) Calculated phonon dispersions of 1\textit{T'}-$\mathrm{MoTe}_2$.  A$_g$-symmetric phonon modes that are coherently excited and observed in the experiment are marked in red.}
        \label{fig1}
\end{figure}

$\mathrm{MoTe}_2$ belongs to the family of TMDCs. Its crystal structure consists of weakly coupled layers held together by van der Waals interactions. At room temperature, $\mathrm{MoTe}_2$ adopts the centrosymmetric, monoclinic 1\textit{T'} phase. Its crystal structure and Brillouin zone are shown in Figs. \ref{fig1}(b) and \ref{fig1}(c), respectively. The left panel of Fig. \ref{fig1}(d) compares experimental ARPES data of 1\textit{T'}-$\mathrm{MoTe}_2$ and the result of a DFT band structure calculation (dashed lines) along the $\Gamma$–X high-symmetry direction. Both experimental data and the results of the band structure calculations agree well with the results of previous work \cite{crepaldi2017persistence,park2025lifshitz}. The right panel of Fig. \ref{fig1}(d) shows the spectral function calculated from the DFT band structure under consideration of electron-phonon interaction in a many-body framework.
Overall, experimental data and calculations agree well, in particular in the binding energy range between the Fermi level $E_\text{F}$ and \SI{-0.4}{eV}. 
We assign differences between experiment and calculations, such as the clear discrepancy in spectral weight near $\Gamma$, to effects such as the coupling to final state re\-so\-nan\-ces or a possible \(k_z\)-offset, at which the \SI{5.9}{eV} photons probe the electronic structure with respect to the $\Gamma$-X-Y plane. In a photoemission study of the related compound $\mathrm{WTe}_2$, we have discussed the origins of these differences in more detail \cite{hein2020mode}. Further differences may arise from surface states and resonances in the ARPES data that are not considered in the first principle calculations \cite{damascelli2004probing}.\\
Figure \ref{fig1}(e) shows the phonon dispersions of 1\textit{T'}-$\mathrm{MoTe}_2$ along the $\Gamma$–X and $\Gamma$–Y directions, calculated using DFPT. The A$_g$-symmetric phonon modes marked in red are relevant for the analysis of the tr-ARPES data. In previous time-resolved all-optical studies, the coherent excitation of these modes was experimentally demonstrated \cite{fukuda2020ultrafast,zhang2019light}.

\begin{figure}
        \includegraphics[width=1\linewidth]{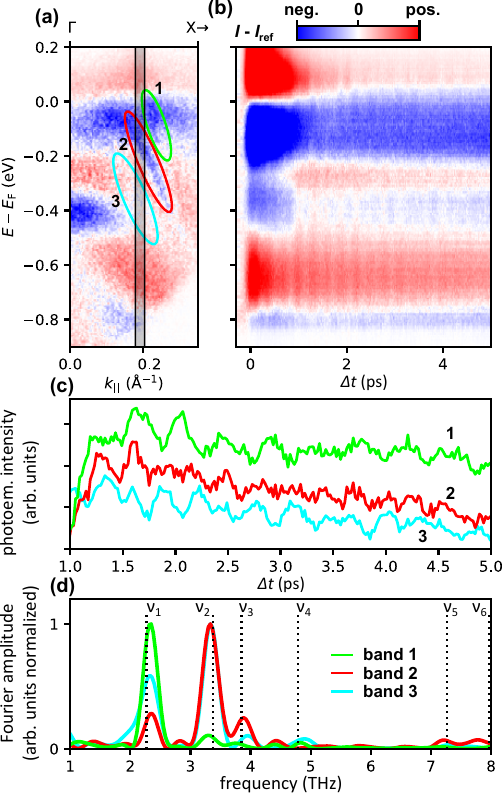}
        \caption{(a) Tr-ARPES spectrum along \(\Gamma\)-X at \(\Delta t=\SI{1.5}{\pico s}\). To highlight the nonequilibrium response to  photoexcitation, a tr-ARPES spectrum recorded at negative \(\Delta t\)  has been subtracted from the raw data. Red (blue) regions denote gain (loss) of spectral weight relative to the equilibrium state. The ROIs marked in green, red, and turquoise indicate  different electronic bands. (b) Difference EDCs at $k_{\parallel}\approx \SI{0.18}{{\angstrom^{-1}}}$ as a function of $\Delta t$. The grey shaded area marked in (a) indicates the momentum range that was used for integration. The oscillations in photoemission hint to the excitation of coherent phonons. (c) Transient integral photoemission intensities for \(\Delta t>\SI{1}{\pico s}\) from the three ROIs marked in (a). (d) Normalized Fourier spectra of the transients in (c), revealing band-specific couplings to different coherent phonon modes.}
    	\label{fig2}
\end{figure}

Figure \ref{fig2}(a) presents tr-ARPES data of 1\textit{T'}-$\mathrm{MoTe}_2$ recorded \SI{1.5}{\pico s} after excitation by the pump pulse. To isolate the transient response of the electronic system upon photoexcitation, a reference spectrum recorded prior to excitation was subtracted from the raw data. Blue regions indicate a loss of spectral weight compared to the state before excitation, while red regions indicate a gain. The areas highlighted in green, red, and turquoise mark spectral regions of three different electronic bands identified in the ARPES spectrum that will be referred to as 'band 1', 'band 2', and 'band 3' in the following. Figure \ref{fig2}(b) shows difference energy distribution curves (EDC) at $k_{\parallel}\approx \SI{0.18}{{\angstrom^{-1}}}$ [see gray marked area in Fig. \ref{fig2}(a)] as a function of pump–probe delay $\Delta t$. The gain in spectral weight above $E_\text{F}$ and part of the initial loss of spectral weight below $E_\text{F}$ indicate photoexcitation of charge carriers due to the absorption of the near-infrared (NIR) pump pulse.

We observe additional changes in the spectral weight distribution that are clear signatures of laser-induced dynamics in the electronic system beyond pure charge carrier dynamics. This applies in particular to the increase in spectral weight in some regions below $E_\text{F}$ and the pronounced time-periodic changes observable on time scales $\Delta t \gtrsim \SI{1}{\pico s}$. Such oscillatory behavior in the electronic structure is a typical indication of the excitation of coherent phonons that couple to the electronic system through strong electron-phonon interaction. It is precisely these changes that will be the focus of the following data analysis and discussion.

Figure \ref{fig2}(c) displays the integral photoemission intensities as a function of $\Delta t$ extracted from the three regions of interest (ROI) marked in Fig. \ref{fig2}(a), each covering the response of a distinct electronic band. In the graph, we have disregarded data for $\Delta t < \SI{1}{\pico s}$, as the transient signal is dominated by the charge carrier dynamics in this time range. All three transients exhibit clear oscillatory behavior. Contributions from different oscillation periods can be seen in the data, with the relative amplitudes varying among the bands.

An analysis using a fast Fourier transform (FFT) allows us to resolve the different frequency contributions to the photoemission intensity transients.
Prior to FFT analysis, the transients were background-corrected using a polynomial fit to remove non-oscillatory components associated with charge carrier population dynamics. Details of the analysis can be found in the supplemental information \cite{Supplemental}. The FFT results are shown in Fig. \ref{fig2}(d). Overall, we observe six spectral peaks at $\nu_1=2.34~\mathrm{THz}$, $\nu_2=3.34~\mathrm{THz}$, $\nu_3=3.86~\mathrm{THz}$, $\nu_4=4.81~\mathrm{THz}$, $\nu_5=7.25~\mathrm{THz}$ and $\nu_6=7.78~\mathrm{THz}$. These frequencies match the frequencies of the six A$_g$-symmetric phonon modes marked in Fig. \ref{fig1}(e) quite reasonably and show very good agreement with the results of other time-resolved studies of $\mathrm{MoTe}_2$ \cite{zhang2016raman,zhang2019light,chen2016activation,beams2016characterization,fukuda2020ultrafast, ma2016raman, lai2018anisotropic}.\\
Each band has a characteristic spectral fingerprint. The data of band 1 shows a dominance of the $\nu_1$ mode, while the amplitudes of $\nu_2$ and $\nu_3$ are greatly reduced in comparison. The signal for band 2 suggests a strong coupling to $\nu_2$ in the selected energy-momentum region, while $\nu_1$ and $\nu_3$ are significantly weaker in amplitude. For this band, we also observe small amplitudes near $\nu_5$ and $\nu_6$. In the case of band 3 $\nu_2$ is again the strongest mode, but the relative coupling to $\nu_1$ is notably enhanced compared to band 2. In addition, the small signal at $\nu_3$ and $\nu_4$ could also indicate weak coupling to these modes.

\begin{figure}
        \includegraphics[width=\linewidth]{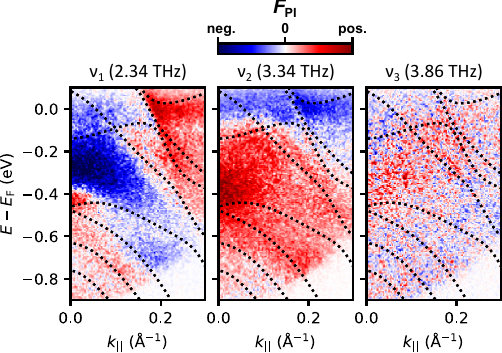}
        \setlength{\abovecaptionskip}{-10pt}
        \caption{FDARPES maps for the three  coherent phonon modes ($\nu_1$, $\nu_2$, $\nu_3$) dominating the electronic response. The data are overlaid with DFT band structures for comparison (dashed lines). The color intensity is a measure of the Fourier amplitude, while the color coding represents its sign, so that both strength and phase of the response of the electronic system to the phonon excitation are depicted.}
    	\label{fig3}
\end{figure}

To further explore the band-selectivity of EPC in 1\textit{T'}-MoTe$_2$, we employ the FDARPES method for data analysis \cite{hein2020mode, Suzuki2021, Lee2023, Ren2023}. In this method, the tr-ARPES data is subjected to a pixel-by-pixel Fourier analysis. This enables an intuitive and phonon mode-specific representation of the electronic response due to EPC.\\
Figure \ref{fig3} compares results of such an analysis for the frequency components $\nu_1$, $\nu_2$, and $\nu_3$, i.e., for the three low-frequency phonon modes that are identified in the tr-ARPES data. Details on data analysis are described in the supplementary information \cite{Supplemental}. The data (in the following referred to as FDARPES maps) show for a selected phonon mode \(\nu_i\) the real part of the Fourier component $F_\text{PI}$ as a function of electron energy \(E\) and wave vector \(k\). Here, the Fourier amplitude is represented by the color intensity, while the relative phase of the photoemission response in the different regions is encoded in the blue-red contrast, which indicates a phase difference of $\pi$ or close to $\pi$.

Comparison of the FDARPES maps reveals pronounced differences in both phase and amplitude patterns implying that the different phonon modes couple very differently to the electronic states, similar to previous results for the related compound $\textit{Td}$-WTe$_2$~\cite{hein2020mode, gauthier2025analysis}. In addition, when compared to the DFT band structure (dashed lines) the electronic bands can in some areas be traced in the FDARPES maps rather well and often separate regions of opposite phase response. The dispersive contours are partly even better visible than in the ARPES spectrum in Fig. 1(d). 

The oscillations in the photoemission intensities that are encoded for each phonon mode in the FDARPES maps via $F_\text{PI}$ can have different origins \cite{gauthier2025analysis}. This includes (i) oscillations in the band energies (band renormalizations), (ii) oscillations of the integral spectral weight, e.g., as a result of changes in the photoemission matrix element, and (iii) oscillations of the band-intrinsic linewidths. In general, these contributions superimpose each other in the Fourier signal. Using an elaborated analysis method described in detail in Ref.~\cite{gauthier2025analysis}, it is possible to partially separate the different contributions in the spectra. In the following, we will apply the analysis procedure to our data that is referred in~\cite{gauthier2025analysis} to as photoemission intensity analysis (PI analysis) focusing specifically on band renormalizations due to coherent phonon excitations. 

\begin{figure*}[]
        \includegraphics[width=1\linewidth]{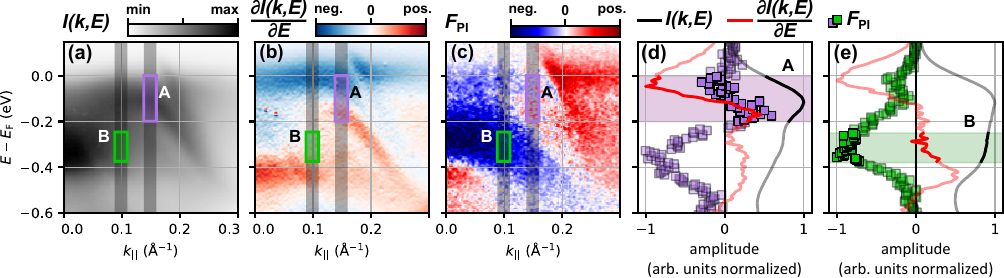}
        \setlength{\abovecaptionskip}{-10pt}
        \caption{(a) Selected energy-momentum region of the ARPES data \(I(k,E)\) shown in Fig. 1(d). (b) Energy gradient $\partial I(k,E)/\partial E$ as a function of energy and momentum of the data shown in (a). (c) FDARPES map for the $\nu_1$ mode in the energy-momentum region covered by the ARPES data in (a). (d), (e) Comparison of \(I(E)\), $\partial I(k,E)/\partial E$, and \(F_\mathrm{PI}\) along the grey shaded areas marked in Figs. (a)--(c). The discussion in the text focuses on  ROI A (d) and B (e) marked in purple and green. All data in (d) and (e) are normalized to their maximum value.}
    	\label{fig4}
\end{figure*}

Contributions to $F_\text{PI}$ due to time-dependent band renormalizations \(\delta E(t)\) can be directly related to the derivative of the static ARPES intensity \(I(k,E)\) with respect to energy \(E\), $\partial I(k,E)/\partial E$. For a sufficiently small renormalization amplitude $\Delta E$, the photoemission intensity can be expanded as \(I(k,E + \delta E(t)) \approx I(k,E) + \delta E(t) \partial_E I(k,E)\), so that, to first order, the Fourier component at the phonon frequency is given by \cite{Giovannini2020}

\begin{equation}
\label{eq:PI_analysis}
    F_\text{PI} \approx -\Delta E \times \frac{\partial I(k,E)}{\partial E}.
\end{equation}

The equation implies that, for an isolated band, this contribution to $F_\text{PI}$ is antisymmetric with respect to the energy band position and has a zero crossing at this energy.

If $F_\text{PI}$ originates from oscillations in spectral weight it can be written as

\begin{equation}
\label{eq:SW}
    F_\text{PI} = \Delta SW \times I(k,E),
\end{equation}

with $\Delta SW$ being the amplitude of the spectral weight oscillation. This contribution to $F_\text{PI}$ is symmetric with respect to the energy band position and shows a maximum at this energy.

There is no similarly simple analytic relation between $F_\text{PI}$ describing an oscillation in the linewidth. However, as is the case of spectral weight oscillations, this contribution is symmetric with respect to the energy band position and shows a maximum at this energy~\cite{gauthier2025analysis}.

The characteristic spectral shapes of the three contributions make it possible to partly distinguish them in the FDARPES maps. Furthermore, Eq.~\ref{eq:PI_analysis} can be used to quantify energy shifts $\Delta E$ due to band renormalizations.

Figures~\ref{fig4}(a)--(c) compare a static ARPES spectrum $I(k,E)$ of 1\textit{T'}-$\mathrm{MoTe}_2$, its derivative $\partial I(k,E)/\partial E$, and the FDARPES map for $\nu_1$ within a selected energy-momentum window.  Furthermore, Figs.~\ref{fig4}(d) and \ref{fig4}(e) compare the normalized momentum-integrated profiles from the purple and green marked ROIs A and B in Figs.~\ref{fig4}(a)--(c).\\
In  ROI A [Fig. \ref{fig4}(d)], we focus on the energy range between $E_\mathrm{F}$ and \(E-E_\mathrm{F}=\SI{-0.2}{eV}\). Here, both \(F_\text{PI}\) and \(\partial I(k,E)/\partial E\) show matching zero crossings, indicating a proportionality according to Eq.~(1) and thus an energy shift \(\Delta E\) due to a band renormalization. The zero crossing simultaneously marks the energy band position (see above) and indeed coincides with a maximum in \(I(k,E)\). This explains why the calculated band dispersions in the FDARPES maps in Fig. \ref{fig3} tend to trace boundaries separating regions of opposite phase.\\
In region B [Fig.~\ref{fig4}(e)], $F_\mathrm{PI}$ shows no zero crossing. Instead, we observe that within the energy range $\SI{-0.35}{eV}\lesssim E-E_\text{F}\lesssim \SI{-0.25}{eV}$ highlighted in green $F_\text{PI}$ shows a distinct maximum that coincides with the low energy maximum in the double peak structure of $I(k,E)$. This indicates dominance of spectral-weight or linewidth oscillations in this region. Outside this window $F_\mathrm{PI}$ cannot be clearly assigned to either $I(k,E)$ or $\partial I(k,E)/\partial E$, suggesting a superposition of different contributions or overlapping bands.\\
Overall, this comparison highlights the complexity of the EPC landscape captured by our FDARPES data of 1\textit{T'}-$\mathrm{MoTe}_2$ and demonstrates that interpreting the origin of the Fourier signal requires a careful, region-specific analysis as outlined in detail in Ref. \cite{gauthier2025analysis}. 

To gain quantitative information on the band energy shifts $\Delta E$ due to band renormalizations, we perform a linear regression between the $F_\text{PI}$ signal and $\partial I(k,E)/\partial E$ using sliding windows across the full FDARPES maps (PI analysis). According to Eq.~\ref{eq:PI_analysis}, the regression slope directly yields $\Delta E$. We performed a PI analysis for the data for \(\nu_1\) and \(\nu_2\). Due to the low overall contrast in the FDARPES map [see Fig. \ref{fig3}(c)] such an analysis did not deliver meaningful results for \(\nu_3\).
A helpful tool supporting the PI analysis is the so-called first-moment analysis (FM analysis). As shown in Ref.~\cite{gauthier2025analysis}, in an FM analysis the influence of spectral weight oscillations are suppressed, allowing band renormalizations to be identified by comparing the energy and momentum dependence of the FM analysis to the results of the PI analysis.  Details of the FM analysis of our data are described in the supplementary information \cite{Supplemental}\\
Additionally we performed calculations using the \textit{ab initio} framework described in the methods section under consideration of the experimental excitation parameters. The calculations generate time- and momentum-resolved spectral functions in which only the phonon-induced band renormalizations are taken into account. To enable the direct comparison, the theory results are finally analyzed in the same way as the experimental data.

\begin{figure*}[]
        \includegraphics[width=17cm]{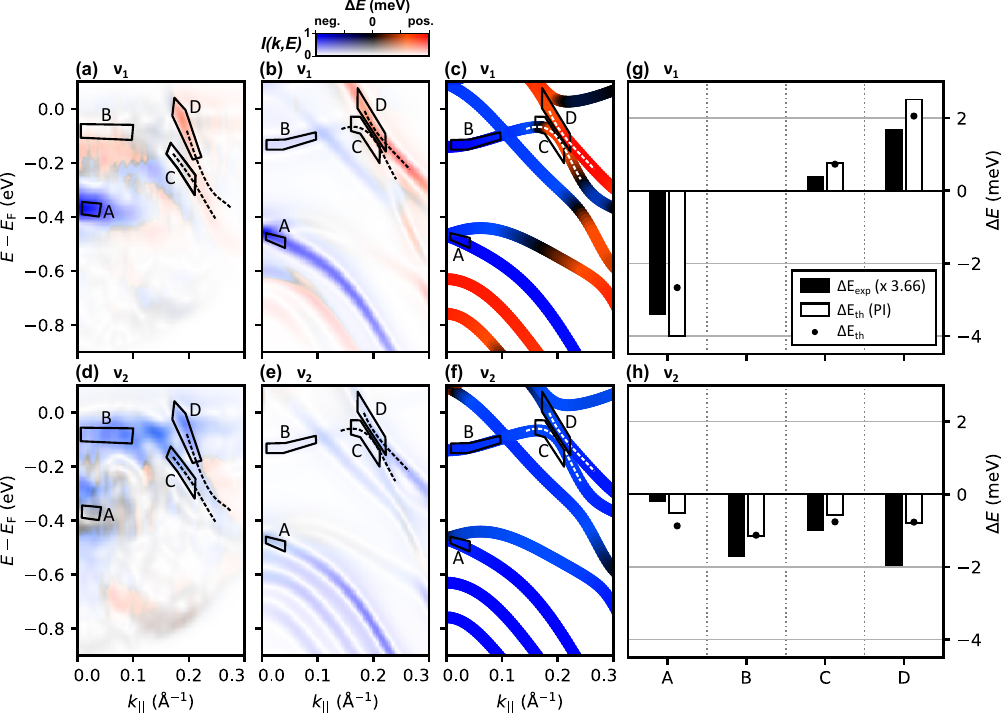}
        \setlength{\abovecaptionskip}{-0pt}
        \caption{(a), (d) Results of the PI analysis of the experimental data for $\nu_1$ (a) and $\nu_2$ (d). (b), (e) Corresponding results of the PI analysis of the theory data for comparison. (c), (f) \textit{ab initio} data of the band renormalization used to generate the data shown in (b) and (e). The dashed lines in the figures indicate an avoided crossing in the band structure as deduced from the band structure mask determined from the experimental FDARPES data [(a), (d)] and the DFT band structure [(b), (c), (e), (f)]. The evaluated energy shifts $\Delta E$ due to band renormalizations are superimposed with a mask derived from an edge detection applied to the corresponding FDARPES maps and the photoemission intensity to highlight only contributions to the data near or in the vicinity of electronic bands. The five ROIs (A--D) marked in (a)--(f) indicate different electronic bands. (g), (h) Amplitudes of \(\Delta E\), determined from the respective average value in the ROIs A--E in (a) and (d) as well as (c) and (f). The experimental values are scaled with the mean ratio between the experimental and theoretical values. Black points show the \textit{ab initio} values of the amplitudes in the ROIs displayed (c) and (f).}
    	\label{fig5}
\end{figure*}

Figure~\ref{fig5} compares the PI-analyzed experimental data [(a), (d)] and the PI-analyzed results of the \textit{ab initio} calculations [(b), (e)] for the two phonon modes $\nu_1$ and $\nu_2$. The direct theoretical band renormalizations from the \textit{ab initio} calculations visualized along the DFT band structure are shown in Figs.~\ref{fig5} (c) and (f).  To isolate the band-specific response in the experimental data, we exploit phase boundaries in the FDARPES maps as band-position markers. An edge detection filter identifies these boundaries, generating masks that restrict the evaluation of $\Delta E$ and its appearance in Figs. 5 (a) and (d) to the band structure detected in the experiment (see~\cite{Supplemental}). The displayed color intensity is a measure of the amplitude of the band renormalization \(\Delta E\), while the color coding represents the sign of \(\Delta E\) at time zero, i.e., the initial direction of the band renormalization.\\
In the experimental data we select ROIs A--D [see markings in Figs.~\ref{fig5}(a) and \ref{fig5}(b)], based on a qualitative agreement of the PI analysis data with the FM analysis data \cite{Supplemental, gauthier2025analysis}. Matching ROIs are marked in the PI-analyzed theory data in Figs.~\ref{fig5}(c) and \ref{fig5}(d). The match was ensured by choosing identical momentum ranges and adjusting the energy ranges according to the differences in the experimental and theoretical band structures.\\

Overall, we observe for both modes a very good qualitative agreement between the PI-analyzed experimental and theory data.  
The $\nu_1$ mode shows particular strong band renormalizations  in ROIs A and D with initial energy shifts in opposite directions. In ROI C, both experiment and theory  hint to a phase reversal at \(\approx\SI{0.19}{\angstrom^{-1}}\). Its origin could stem from an avoided crossing of two bands in this region, indicated by the dashed lines. The apparent discrepancy between experiment and theory in the initial sign of the band renormalization in ROI B appears to be due to a signal contribution in the experimental data from the high binding energy region below ROI B. We associate the finite signal amplitude in this energy region with $\partial I(k,E)/\partial E$ being very close to zero [cf. Fig. \ref{fig4} (b)]. This leads to a significant and erroneous signal enhancement according to equation (1). We assume that this signal, which we cannot be related to a band, is an artifact of the analysis procedure. It will not be considered in the further discussion.

For the $\nu_2$ mode, experiment and theory show for all four ROIs a good agreement regarding the initial sign of band renormalization and the relative renormalization amplitudes. Furthermore, experiment and theory also agree qualitatively regarding the comparison of the relative amplitudes between $\nu_1$ and $\nu_2$. \\

For a quantitative comparison, we determined the renormalization amplitudes $\Delta E$ for ROIs A--D by signal averaging over the respective ROI. The experimental and calculated values for $\Delta E$ are of the same order of magnitude, but the experiment systematically yields smaller amplitudes than the theory. Experimental renormalization amplitudes range between \SI{0.05}{meV} and \SI{0.9}{meV}. Very similar values have been reported for similar excitation conditions for the related compound $\mathrm{WTe}_2$ ~\cite{gauthier2025analysis}. Theory yields renormalization amplitudes between \SI{0.5}{meV} and \SI{4.0}{meV}. 
 
A direct comparison of the renormalization amplitudes for ROI A--D is shown in Fig.~\ref{fig5}(e).
Here, the experimental data are scaled by a factor of 3.66, the average of the ratio between calculated and experimental $\Delta E$ for the  four ROIs. In addition to the theory values extracted from the PI-analyzed data, we also added the actual \textit{ab initio} values for $\Delta E$ [(Fig.~\ref{fig5}(c) and (f)] to the graph (black dots). This data representation confirms the very good match between experiment and theory regarding the relative amplitudes of the band renormalizations in the different ROIs and their initial sign. 
Interestingly, we observe a substantial quantitative deviation between the theory value derived from the PI-analyzed data and the actual \textit{ab initio} value for ROI A. According to theory, in this area two bands overlap. This indicates that in such a scenario, a PI-analysis yields quantitatively inaccurate values in accordance with findings reported in Ref. \cite{gauthier2025analysis}.  

The systematic deviation between experimental and calculated renormalization amplitudes can have various causes: The tendency of a systematic underestimation of excitation amplitudes by the evaluation method as reported in \cite{gauthier2025analysis}, the possible \(k_\mathrm{z}\)-offset in the experimental photoemission data with respect to the \(\Gamma\)-X-Y plane considered in the theory, and typical uncertainties in determining absorbed excitation fluence. Finally, residual spectral-weight or linewidth contributions to the experimental data cannot be completely excluded.

\section*{Conclusion}

In summary, we have demonstrated the band-selectivity of coherent phonon-driven band renormalizations in the semimetal 1\textit{T}$'$-MoTe$_2$. The frequency-domain ARPES (FDARPES) method reveals insights into the mode-selective response of the electronic structure to the excitation of coherent phonons.\\
The application of a recently introduced PI analysis method delivers quantitative results on amplitude and initial direction of coherent phonon-driven band renormalizations, while the FM analysis is used to strengthen the robustness of the interpretation \cite{gauthier2025analysis}. For an excitation density in the few \(\SI{100}{\micro J/cm^2}\) range we observe in the experiment typical renormalization amplitudes of a few \SI{100}{\micro eV} for different bands and  the two phonon modes at frequencies $\nu_1=\SI{2.34}{THz}$ and $\nu_2=\SI{3.34}{THz}$ that dominate the electronic structure response in this material.\\
\textit{Ab initio} calculations reproduce the initial directions of band renormalization for both $\nu_1$ and $\nu_2$ remarkably well. The calculated amplitudes of band renormalizations are of the same order of magnitude as in the experiment, but systematically higher. The overall very good qualitative agreement between theory and experiment suggests that such work will enable further insights into electron-phonon coupling phenomena in quantum materials in the future, thanks to recent developments in the methods used to evaluate such experimental data \cite{hein2020mode, gauthier2025analysis} and the parallel advances made in the theoretical description of these phenomena \cite{emeis2025coherent}.

\begin{acknowledgments}
This work is funded by the Deutsche Forschungsgemeinschaft (DFG), Projects No. 443988403 and No.
499426961. The authors gratefully acknowledge the computing time provided by the high-performance computer Lichtenberg at the NHR Centers NHR4CES at TU Darmstadt (Project p0021280).
\end{acknowledgments}

\section*{Data Availability}

The experimental data that support the findings of this study are openly available on Zenodo at http://doi.org/10.5281/zenodo.18671857.

The theory data that support the findings of this study are openly available in the NOMAD database at http://doi.org/10.17172/NOMAD/2026.02.19-3.

\section*{Author Declarations}

\subsection*{Conflict of Interest}
The authors have no conflicts to disclose.

\subsection*{Author Contributions}
Carl E. Jensen and Michael Bauer conceived the experiments. Carl E. Jensen and Stephan Jauernik carried out the experiments. Carl E. Jensen and Petra Hein performed the data analysis. Christoph Emeis performed the calculations. All authors contributed to the scientific discussion. Carl E. Jensen, Christoph Emeis, Fabio Caruso, and Michael Bauer cowrote the paper.

\nocite{SobelFilter,caruso2022ultrafast,pan2023vibrational,PhysRevB.107.054102,RevModPhys.89.015003,dal2014pseudopotentials,PhysRevLett.100.136406,PhysRevB.13.5188,bosoni2024verify,brown1966crystal,baroni2001phonons,marzari2012maximally,pizzi2020wannier90}

\bibliography{sources}

\end{document}


\title{Supplementary Materials for ``Coherent Phonon-Driven Band Renormalizations in 1\textit{T}$'$-MoTe$_2$''}
\author{Carl E.~Jensen}
\email{cjensen@physik.uni-kiel.de}
\affiliation
{Institute of Experimental and Applied Physics, Kiel University, 24118 Kiel, Germany
}
\author{Christoph~Emeis}
\affiliation{Institute of Theoretical Physics and Astrophysics, Kiel University, 24118 Kiel, Germany
}
\author{Stephan~Jauernik}
\affiliation{Institute of Experimental and Applied Physics, Kiel University, 24118 Kiel, Germany
}
\author{Petra~Hein}
\affiliation{Institute of Experimental and Applied Physics, Kiel University, 24118 Kiel, Germany
}
\author{Fabio~Caruso}
\affiliation{Institute of Theoretical Physics and Astrophysics, Kiel University, 24118 Kiel, Germany
}
\author{Michael~Bauer}
\homepage{https://www.physik.uni-kiel.de/de/institute/ag-bauer}
\affiliation{Institute of Experimental and Applied Physics, Kiel University, 24118 Kiel, Germany
}
\affiliation{
Kiel Nano, Surface and Interface Science KiNSIS, Kiel University, 24118 Kiel, Germany
}
\date{\today}

\maketitle

\section{Generation of FDARPES maps}
\label{sec:SI_FDARPES}

FDARPES maps are generated from the tr-ARPES data as follows: For \(\Delta t  \lessapprox 800~\mathrm{fs}\) the transient changes observed in the tr-ARPES data are dominated largely by charge carrier dynamics. To isolate the oscillatory component of the signal from this signal contribution we omit all data for time delays $\Delta t < 810~\mathrm{fs}$ for further evaluation. In the following, the transient photoemission signal from each pixel of the tr-ARPES maps is considered separately. First, fifth-order polynomials are fitted to and subtracted from the transients to account for the low-frequency background of the data. Next, these transients are zero-padded to three times of their original length to increase the frequency resolution of the Fourier signal. The resulting transients are Fourier transformed and compiled into FDARPES maps, which provide the energy- and momentum-dependence of the Fourier signal for each frequency. In a final step, the Fourier signals are scaled by a factor of  $\frac{2}{N}$ with N being the number of sampling values. This procedure restores the correct magnitude of the oscillation amplitude, which is crucial for the further quantitative analysis, such as the PI analysis.

\section{Determining a global time zero}
\label{sec:SI_TZ}
\begin{figure}
    \centering
    \includegraphics[width=1\linewidth]{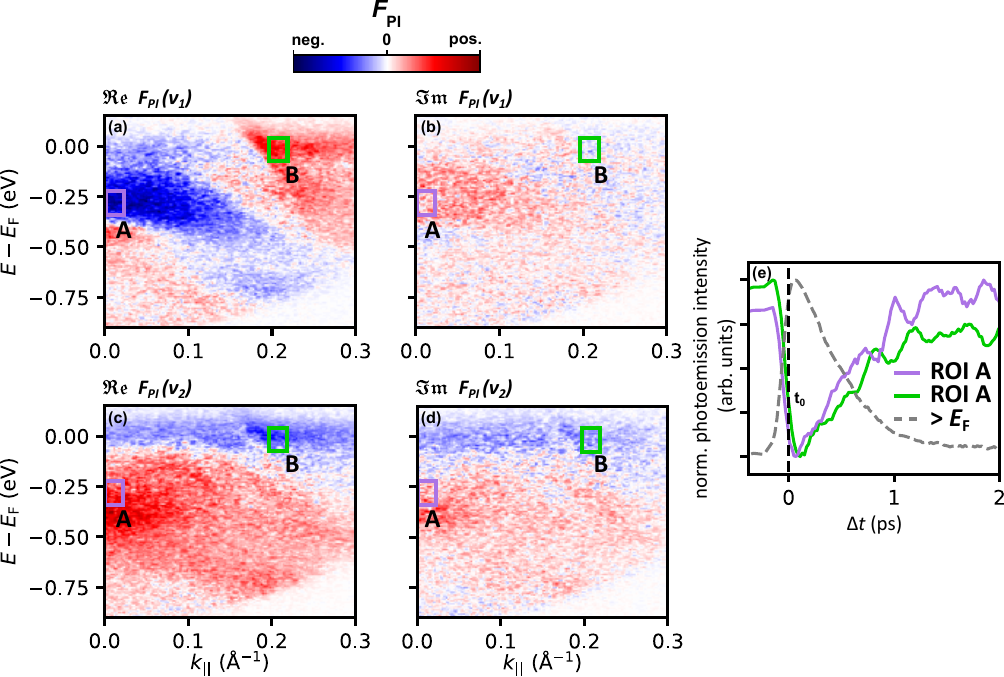}
    \caption{(a)--(d) Real ($\Re$) and imaginary ($\Im$) part of $F_\text{PI}(k,E)$ for $\nu_1$ (a,b) and $\nu_2$ (c,d). (e) Transient photoemission signal from  ROIs A and B in comparison to a photoemission intensity transient  at $E-E_\mathrm{F}\approx $ representing the temporal evolution of the photoexcited nascient electron population. }
    \label{fig:TimeZero}
\end{figure}

For the PI and the FM analysis, an effective and common time zero $t_0$ for the launch of $\nu_1$ and $\nu_2$ must be determined. Two boundary conditions must be observed: i) Either imaginary part or real part of the Fourier components $F_\text{PI}(k,E)$ must be minimized simoultaneously for $\nu_1$ and $\nu_2$ \cite{gauthier2025analysis}. ii) $t_0$ must lie within the initial rise of the nascent carrier population generated in the photoabsorption process.\\
For our analysis we chose $t_0$ so that the imaginary part of $F_\text{PI}(k,E)$ is minimized. Fig.~\ref{fig:TimeZero} compares the real part [(a) and (c)] and imaginary part [(b) and (d)] of $F_\text{PI}(k,E)$ at $t_0$ for $\nu_1$ and $\nu_2$, respectively. We suspect that the limited time resolution of the experiment (\SI{130}{\femto s}) and an incomplete subtraction of signal background are responsible for the residual signal visible in the imaginary part of the Fourier component. 
Fig.~\ref{fig:TimeZero} (e) shows the transient photoemission signal from the two regions of interest (ROI) A and B marked in Fig.~\ref{fig:TimeZero} (a)--(d) in comparison to the charge carrier population dynamics at an excitation energy $E-E_\mathrm{F}\approx $ (grey dashed line), confirming that $t_0$ lies within the initial rise of the photoexcited nascent electron population.

\section{Generation of Band Structure Masks}
\label{sec:SI_MASK}

\begin{figure}
    \centering
    \includegraphics[width=0.95\linewidth]{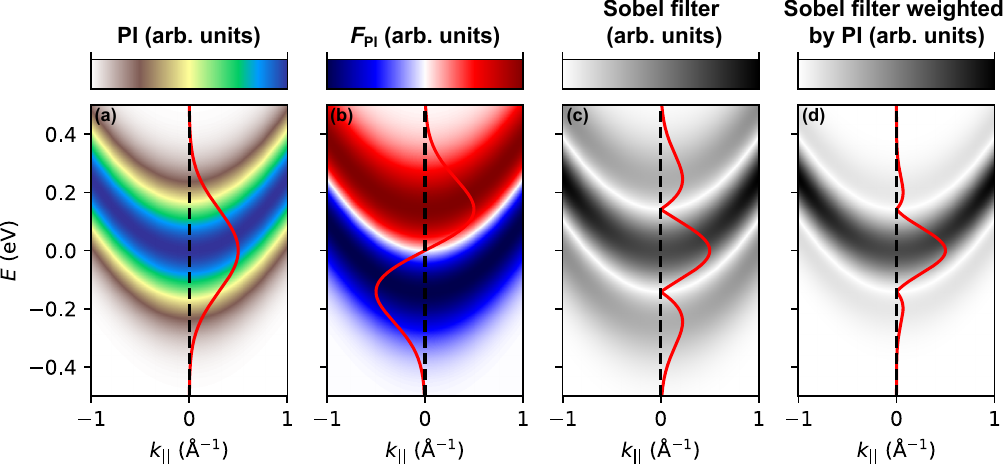}
    \caption{(a) Example of an electronic band to illustrate the method used to generate a band structure mask. For the following analysis the band energy is periodically modulated with an amplitude of \SI{0.1}{eV}. (b) FDARPES map of the of data shown in (a) for the chosen modulation frequency. (c) FDARPES map after application of a Sobel edge detection filter. (d) Band structure mask generated from multiplying Fig. (c) with the normalized photoemission intensity shown in (a). The red lines in  (a)--(d) show the respective signal amplitude along the black dashed line.}
    \label{fig:edge_detect}
\end{figure}

\begin{figure}
    \centering
    \includegraphics[width=0.8\linewidth]{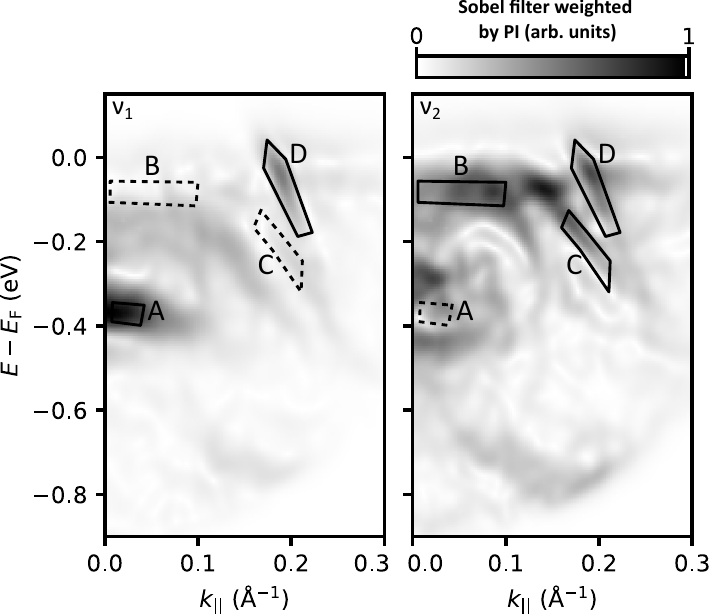}
    \caption{Band structure masks created by applying the described edge detection method to the experimental FDARPES data for $\nu_1$ and $\nu_2$. ROIs from mentioned in the main text are marked.}
    \label{fig:edge_masks}
\end{figure}

To correlate the evaluated band renormalizations with the electronic band structure, it is important to determine the band positions based on the experimental photoemission data. In Ref.  \cite{gauthier2025analysis} this is achieved by determining the two-dimensional curvature of the photoemission intensity. Here, we take a different approach and use a fingerprint of the electronic band positions in the FDARPES maps \cite{hein2020mode}. Fig.~\ref{fig:edge_detect}(a) shows the simulation of an example electronic band with parabolic dispersion. To account for coupling to a coherent phonon mode, the band position is periodically modulated with an amplitude \SI{0.1}{eV} in energy direction. The corresponding FDARPES map is shown in Fig.~\ref{fig:edge_detect}(b). The actual band position is captured in this data by the zero crossing of the Fourier signal separating the blue and red regions. Such boundary lines are also clearly visible in the experimental data in Fig. 3. A filter based on this signature can therefore further increases the sensitivity to band renormalization effects. 
To generate a band structure mask for the analysis of band renormalizations, we track this feature by applying a Sobel filter to the FDARPES map \cite{SobelFilter}. To avoid artifacts from signal noise in the experimental data, we apply a Gaussian blur before filtering.\\
The filtered image of Fig.~\ref{fig:edge_detect}(b) is shown in Fig.~\ref{fig:edge_detect}(c). The red line shows the intensity profile  along the dashed line. The intensity is maximum at the actual band position, but secondary maxima appear above and below it. To suppress these artifacts, we multiply the filtered image with the photoemission intensity of the ARPES spectrum, resulting in the band structure mask shown in Fig.~\ref{fig:edge_detect}(d). Fig. \ref{fig:edge_masks} shows the masks used to generate Figs. 5(a) and 5(b) in the main text, calculated from the experimental FDARPES data for the phonon modes $\nu_1$ and $\nu_2$. The ROIs from the main text are marked. Solid lines indicate that the ROI is chosen on the basis of the respective mask.

\section{First moment (FM) analysis}
\label{sec:SI_FM}

To quantify band renormalizations from the experimental data, we apply the photoemission intensity (PI) analysis method to the experimental data, which is described in detail in Ref. \cite{gauthier2025analysis}. An alternative method presented in the same study is based on a first moment (FM) analysis of the energy distribution curves of the tr-ARPES spectra. The quantitative validity of this method is less reliable, however, it is sensitive exclusively to band renormalization effects. It can therefore be used as a benchmark to clearly identify band renormalizations in the PI analysis data. The FM of an energy distribution curve \(I\) is given by \cite{gauthier2025analysis}:

\begin{equation}
    \mathrm{FM}(k,E,t) = \frac{\int^{E+\delta/2}_{E-\delta/2} I(E',k,t) E'dE'}{\int^{E+\delta/2}_{E-\delta/2} I(E',k,t)dE'}
\end{equation}

\begin{figure}
    \centering
    \includegraphics[width=0.75\linewidth]{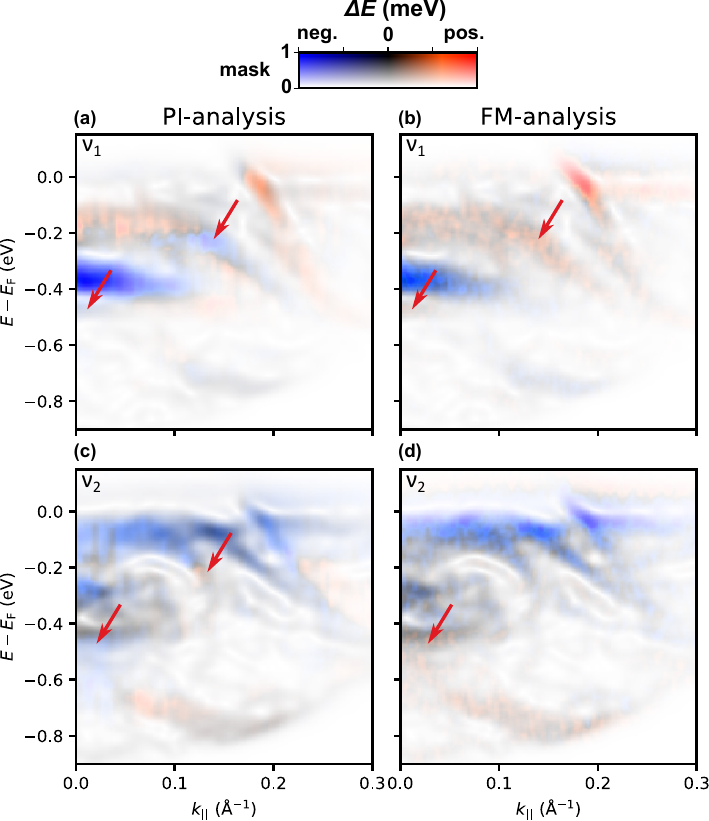}
    \caption{Comparison between the results of the PI analysis [(a) and (c)] and FM analysis [(b) and (d)] of the experimental data for $\nu_1$ and $\nu_2$. The band renormalizations are superimposed on the band structure masks shown in Fig. \ref{fig:edge_masks}. The arrows indicate areas, where PI analysis and FM analysis show opposite signs for \(\Delta E\).}
    \label{fig:SI_FM}
\end{figure}

It is calculated pixel-by-pixel using a sliding window function with the window size $\delta = \SI{0.2}{eV}$. For the following Fourier transform, the time-dependent incoherent background is subtracted by fitting a fifth-order polynomial to the transient FM of each pixel. The size of the Fourier components directly reflect the band renormalization amplitude in this analysis.\\
Fig.~\ref{fig:SI_FM} compares the results of the PI analysis and the FM analysis of the experimental data for $\nu_1$ and $\nu_2$. Over a large part of the spectrum and for both phonon frequencies we observe a very good qualitative agreement between the results of the two analysis methods with respect to the sign of \(\Delta E\). Relevant deviations can only be observed in two areas, which are marked by arrows. At a binding energy of $E-E_{\mathrm{F}}\approx\SI{-0.45}{eV}$ and near $\Gamma$, the PI and FM data show an opposite sign of \(\Delta E\) for both $\nu_1$ and $\nu_2$. At $E-E_{\mathrm{F}}\approx\SI{-0.25}{eV}$ and $k_{||}\approx\SI{0.15}{\angstrom^{-1}}$ we observe an analogous difference for $\nu_1$. The origin for these deviations are most likely contributions from spectral weight or linewidth oscillations, which only affect the result of the PI analysis. These spectral regions are therefore not taken into account for the comparison of experimental and calculated band renormalizations.  

\section{Theory and computational methods}
\label{sec:SI_THEO}

\subsection{Theoretical Framework}
For ultrafast dynamic simulations, the time-dependent Boltzmann equations, as recently implemented in {\tt EPW}, are used to calculate the time-dependent electronic occupations $f_{n\mathbf{k}}(t)$ using the \textit{ab initio} collision integrals for electron and phonon occupations: \cite{caruso2022ultrafast,pan2023vibrational}
\begin{align}
    \label{eq:Gamma1}
    \partial f_{n\mathbf{k}} &= \Gamma^{\rm e-ph}[f_{n\mathbf{k}},n_{\mathbf{q}\nu}] + \Gamma^{\rm e-e}[f_{n\mathbf{k}}] \quad, \\
    \partial n_{\mathbf{q}\nu} &= \Gamma^{\rm ph-e}[f_{n\mathbf{k}},n_{\mathbf{q}\nu}] \quad,
    \label{eq:Gamma2}
\end{align}
where $n_{\mathbf{q}\nu}$ denotes the phonon distribution function, and $\Gamma^{\rm e-ph}$, $\Gamma^{\rm e-e}$, and $\Gamma^{\rm ph-e}$ the collision integrals due to electron-phonon, electron-electron, and phonon-electron scattering, respectively. The explicit expressions for the collision integrals can be found elsewhere. \cite{caruso2022ultrafast}\\

The time-dependent displacement amplitude $Q_{\nu}$ of the coherent phonon mode ${\nu}$ can be obtained by the solution of the coherent-phonon equation of motion: \cite{PhysRevB.107.054102}
\begin{align}
 \partial_t^2 Q_{\nu} + \omega^2_{\nu} Q_{ \nu} =  - \frac{2\omega_{\nu}}{\hbar} \sum_{n \mathbf{k}} g_{nn}^{\nu} (\mathbf{k},0)  \Delta f_{n\mathbf{k}}(t) \quad.
 \end{align}
Here, $\omega_{\nu}$ is the frequency of the coherent-phonon mode and $\Delta f_{n\mathbf{k}}(t) = f_{n\mathbf{k}}(t) - f_{n\mathbf{k}}^{\rm (0)}$ is the change of the electronic occupations compared to the carrier distribution before photoexcitation $f_{n\mathbf{k}}^{\rm (0)}$. 
The band renormalization mediated by the coherent phonons is given by:\cite{emeis2025coherent}
\begin{align}
     \Sigma^p_{n\mathbf{k}}(t) = \sum_{\nu} g_{nn}^{\nu}(\mathbf{k},0)  Q_{\nu}(t)  \quad.
\end{align}
The EPC matrix mediates the change of the eigenenergy that the nuclear motion imposes on the electronic system. The strength of our \textit{ab initio} method lies in its ability to disentangle the contributions of each coherent-phonon mode to the momentum- and energy-resolved band renormalization individually, (as well as the momentum- and energy-resolved electronic contributions to the force acting on the atoms after photoexcitation).

The displacement amplitude can also be linked to the actual nuclear displacement by: \cite{RevModPhys.89.015003}
\begin{align*}
        \Delta {\tau}_{\kappa p} = \sum_{{\bf q}\nu} \left(\frac{\hbar}{ 2
\omega_{\mathbf{q} \nu} M_\kappa}\right)^{\frac{1}{2}} e^{i {\bf q \cdot R}_p}
\, {\bf e}^{\kappa}_{{\bf q}\nu} {Q}_{{\bf q}\nu} \quad, 
\end{align*}
where $\kappa$ and $p$ are the indices of the nucleus and unit cell, $M_\kappa$ is the nuclear mass, ${\bf e}^{\kappa}_{{\bf q}\nu}$ is the phonon eigenvector and ${\bf R}_p$ is the crystal-lattice vector. 

\subsection{Computational details}

Density functional theory (DFT) calculations are performed with the plane-wave pseudopotential code {\tt Quantum ESPRESSO}  \cite{giannozzi2009quantum,giannozzi2017advanced}, using fully-relativistic projector augmented wave (PAW) pseudopotentials \cite{dal2014pseudopotentials} and the Perdew-Burke-Ernzerhof generalized-gradient approximation in solids (PBEsol) for the exchange-correlation functional \cite{PhysRevLett.100.136406}. The plane-wave kinetic-energy cutoff is set to 80~Ry, and the Brillouin zone (BZ) is sampled with a $10\times 6 \times 4$ Monkhorst-Pack grid \cite{PhysRevB.13.5188}. For the BZ integration, we used a Gaussian smearing with a spreading of 0.02\,Ry to mimic an elevated temperature needed for the 1\textit{T’} phase of MoTe$_2$ to emerge \cite{bosoni2024verify}.
The lattice parameter of the monoclinic unit cell with the space group $P2_1/m$ determined through crystal structure and unit cell relaxation are $a = 3.457$, $b = 6.320$, $c = 13.694$ and $\beta = 91.8°$, which agrees well with the experimental values \cite{brown1966crystal}. 
The phonon dispersion is obtained from density functional perturbation theory (DFPT) on a $5 \times 3 \times 2$ q-point mesh \cite{baroni2001phonons}. 
The electron and phonon eigenvalues are interpolated and the electron-phonon coupling matrix $g_{mn}^{\nu}(\mathbf{k},\mathbf{q})$ is calculated on a dense $30\times 18 \times 12$ grid via maximally-localized Wannier functions \cite{marzari2012maximally} within the {\tt EPW} code \cite{lee2023electron}, which uses {\tt Wannier90} as a library \cite{pizzi2020wannier90}.\\

The joint time-dependent Boltzmann equations and the coherent-phonon equation of motion are propagated within an energy window of 1.5\,eV around the Fermi energy. The time propagation is achieved using the second-order Runge-Kutta method over a total simulation time of 6\,ps with a time stepping of 1\,fs. The collision integrals (Eqs.~\ref{eq:Gamma1}-\ref{eq:Gamma2}) are evaluated on the dense $30\times 18 \times 12$ grid with a gaussian smearing of 10\,meV. The lattice temperature is set to room temperature $T_{\rm lat}=300\,$K and the electronic temperature $T_e$ corresponding to the absorbed fluence of the experiment is estimated to be $T_e=1325\,$K according to the procedure described in \cite{emeis2025coherent}. For the electron-electron scattering, an average electron linewidth of $10~$meV for the states within the energy window was chosen. For realistic comparison with the measurements, the experimental energy- and time-resolution were accounted for via Gaussian convolutions.

\clearpage
\bibliography{sources}